# An Auto-tuning Method for Run-time Data Transformation for Sparse Matrix-Vector Multiplication

TAKAHIRO KATAGIRI† MASAHIKO SATO††

In this paper, we research the run-time sparse matrix data transformation from Compressed Row Storage (CRS) to Coordinate (COO) storage and an ELL (ELLPACK/ITPACK) format with OpenMP parallelization for sparse matrix-vector multiplication (SpMV). We propose an auto-tuning (AT) method by using the $D_{mat}{}^i$ - $R_{ell}{}^i$ graph, which plots the derivation/average for the number of non-zero elements per row ($D_{mat}{}^i$) and the ratio, SpMV speedups/transformation time from the CRS to ELL ($R_{ell}{}^i$). The experimental results show the ELL format is very effective in the Earth Simulator 2. The speedup factor of 151 with the ELL-Row inner-parallelized format is obtained. The transformation overhead is also very small, such as 0.01 to 1.0 SpMV time with the CRS format. In addition, the $D_{mat}{}^i$ - $R_{ell}{}^i$ graph can be modeled for the effectiveness of transformation according to the $D_{mat}{}^i$ value.

## 1. Introduction

Sparse matrix-vector multiplication (SpMV) is an important process for scientific computing. However, to obtain higher performance, optimization is needed. First, SpMV needs to optimize computations by taking into account the structure of the sparse matrix. At the numerical library level, an input matrix cannot be obtained until a library call is executed. Hence, run-time optimization is needed. To optimize a computation with respect to input data, auto-tuning (AT) technologies have been proposed and implemented in several numerical libraries[1]-[3]. In addition, SpMV needs dedicated implementations to establish high performance. Many previous studies for automatic code generation have been conducted[4]-[6]. In recent studies, the target areas of AT have been extended to whole levels of the iterative algorithm[7].

In this paper, we focus on the run-time sparse data transformation from Compressed Row Storage (CRS). The CRS format is a very easy and widely used data format in application software, but optimization can be further developed. Although several previous studies focused on the best storage for sparse matrices, including register and cache blocking [7], the data transformation time did not take into account the optimization time. To extend the adaptability of AT to many computer architectures and situations, run-time data transformation cannot be avoided.

This paper is organized as follows. Section 2 explains the sparse data transformation from the CRS format to other formats. In addition, an auto-tuning method for run-time data transformation is proposed. Section 3 shows the parallel implementation with OpenMP by using the COO (Coordinate) format and the ELL (ELLPACK/ITPACK) format. Section 4 is a performance evaluation with the HITACHI SR16000/VL1 and the Earth Simulator 2 (ES2). Finally, we summarize our findings in Section 5.

## 2. Transformation To COO and ELL Formats and A Run-time Auto-Tuning Strategy

### 2.1 Transformation from CRS to COO and ELL Formats Explanation

**Explanation of Sparse Formats**

Let $A \in \Re^{n \times n}$ be a sparse matrix. The CRS format is defined as the following three arrays: VAL(1:nnz), ICOL(1:nnz), IRP(1:n+1), where nnz is the number of non-zero-elements. VAL() stores the values of non-zero elements, ICOL() stores the column indexes, and IRP() stores the row index pointers of the sparse matrix.

The COO format is a natural format for representing the sparse matrix. The COO format is defined as follows: VAL(1:nnz), ICOL(1:nnz), and IROW(1:nnz). VAL() and ICOL() are the same arrays used in the CRS format. IROW() instead of IRP() stores the row indexes. Compared to the CRS format, the COO formant requires much memory space to store the index information.

The ELL format is based on a constant bandwidth of the sparse matrix. The ELL format is defined as follows: VAL(1:n,1:nz), ICOL(1:nnz), where nz is the bandwidth length. If the sparse matrix does not have the exact bands, the value of zero is inserted in the position of missing band parts. Hence, the ELL format requires additional computation, if the matrix does not have the exact number of band elements. Moreover, it requires much memory space.

**Transformation from the CRS format**

† Supercomputing Research Division, Information Technology Center, The University of Tokyo
††Fusion Theory and Simulation Research Division, National Institute for Fusion Science

Transformation from the CRS to the COO or the ELL format is easy if the COO or the ELL format requires row-wise storage. This is because the first CRS column index in each row is known via the row pointer arrays of ICOL().

If the COO or the ELL format requires column-wise storage, the transformation is difficult. In this case, we carry out the following two stages: (1) Phase I: transform CRS to CCS (Compressed Colum Format); (2) Phase II: transform CCS to COO with column-wise storage. In Phase II, the transformation is easy since we know the first row index in each column via the pointer arrays of IROW(). The bottleneck is the transformation from CRS to CCS, and the solution is knowing the number of non-zero elements in each column. But CRS does not have such explicit information. To solve this problem, we first count the number of non-zero elements per column. The code is shown as follows.

```
c   ===Count The Number of Non-zero columns.
    DO I=1, N
      NC_IRP(I) = 0
    ENDDO
    DO I=1, N
      DO J_PTR=IRP(I),IRP(I+1)-1
        II = ICOL(J_PTR)
        NC_IRP(II) = NC_IRP(II) + 1
      END DO
    ENDDO
c   ===Set IRP.
    IRP_T(1) = 1
    DO I=2, N
      IRP_T(I) = IRP_T(I-1) + NC_IRP(I-1)
    ENDDO
    DO I=1, N
      NC_IRP(I) = IRP_T(I)
    ENDDO
c   ===Set Column Numbers.
    DO I=1,N
      DO J_PTR=IRP(I),IRP(I+1)-1
        II = ICOL(J_PTR)
        K = NC_IRP(II)
        NC_IRP(II) = NC_IRP(II)+1
        VAL_T(K) = VAL(J_PTR)
```

```
        ICOL_T(K) = I
      END DO
    END DO
c   ===Copy Back
    DO I=1, N
      IRP(I) = IRP_T(I)
    ENDDO
    IRP(N+1) = NNZ+1
    DO I=1, NNZ
      ICOL(I) = ICOL_T(I)
      VAL(I) = VAL_T(I)
    END DO
```

## 2.2   A Run-time Auto-tuning Strategy

Let $t_{crs}$ be the execution time of SpMV with CRS in a sparse matrix. Let $t_{ell}$ be the execution time of SpMV with the ELL format in the same sparse matrix of CRS. We set a ratio using $t_{crs}$, as follows:

$$SP_{crs/ell} = t_{crs} / t_{ell}. \tag{1}$$

Let $t_{trans}$ be the transformation time from CRS to ELL. We set a ratio using $t_{crs}$ for the transformation, as follows:

$$TT_{ell} = t_{crs} / t_{trans}. \tag{2}$$

With (1) and (2), we define the cost of speedup, including the transformation time from CRS to ELL, as follows:

$$R_{ell} = SP_{crs/ell} / TT_{ell}. \tag{3}$$

According to (3), the cost of 1.0 is defined when we establish a 10x speedup of SpMV in ELL to that of CRS, and if and only if the transformation time to SpMV in CRS is 10.

The ELL format is suitable for uniform distribution of the number of non-zero elements per row. By taking into account this feature, the deviation value for the number of non-zero elements per row is the norm of an application using the ELL format. Let $\mu$ be the arithmetic mean for the number of non-zero elements per row. Let $\sigma$ be the standard deviation of the number of non-zero elements per row in the sparse matrix. By using these parameters, the deviation of non-zero elements per row is defined as:

$$D_{mat} = \sigma / \mu \ . \tag{4}$$

In (3), $R_{ell}$ depends on the computer architecture. In contrast, $D_{mat}$ depends on the distribution of non-zero elements of the sparse matrices, not the computer architecture. By using this characteristic, we propose an off-line and an on-line auto-tuning strategy for run-time sparse matrix transformation.

**The Run-time AT Method**

The run-time AT method has two phases, off-line and on-line.

The off-line phase is done when the library is installed into a new computer architecture. The on-line phase is done in every library call for the data created in the off-line phase. The details are shown as follows.

- Off-line Phase
  (1) Pick up the benchmark sparse matrices. Let $mat_i$ ($i=1,...,m$) be the matrices.
  (2) For $i=1,...,m$
    Measure $t_{crs}^i$ as the time for $mat_i$.
    Measure $t_{trans}^i$ as the time for $mat_i$.
    Compute $R_{ell}^i$ with $t_{crs}^i$ and $t_{trans}^i$ for $mat_i$.
    Compute $D_{mat}^i$ for $mat_i$.
    Endfor;
  (3) Plot a 2-D figure with $D_{mat}^i$ as the X-axis and $R_{ell}^i$ as the Y-axis. The graph is called the **$D_{mat}^i$ - $R_{ell}^i$ graph**.
  (4) Find the largest point of the X-axis such that $R_{ell}^i >= c$ for $i=1,...,m$. This point of the X-axis is denoted as $D^*$. We set **c=1.0 as usual**.
- On-line Phase
  (1) Compute $D_{mat}$ for the input matrix.
  (2) If $D_{mat} < D^*$ then use SpMV with the ELL format. Otherwise, use SpMV with CRS.

**Discussion for the AT Method**

One drawback of the AT method is the enormous iteration time needed to take advantage of the transformation effect. For example, in the case that 1000x speedup is established by SpMV with CRS, 1000 times SpMV is needed to transform to the ELL format. In this case, c=1.0 is established. However, a 1000x speedup is unrealistic using the ELL format in current computer architectures. Our target is 2x–100x speedups to SpMV with CRS. (See the performance evaluation in Section 4.) Hence, the iteration time based on the AT algorithm is approximately 2–100 times. This range is achievable for many iterative solvers.

Another drawback is the additional memory required to adapt the run-time transformation. Approximately 2x or more of memory space is needed in comparison with using CRS. To solve this memory problem, we proposed the "auto-tuning policy" for memory space from user requirements of AT[7]).

## 3. OpenMP Implementations of SpMV

### 3.1 Outer Loop Parallelization with COO-Column Format

If the sparse data is in the column-wise COO format (COO-Column), parallel implementation with OpenMP is as shown in Fig. 1.

In Fig. 1, the outer loop of SpMV is divided according to the number of threads. The region of the loop index is stored in ISTART(K) and IEND(K), where K is the thread identifier. Lines <12>–<16> indicate that a reduction of each computation results from each thread. The reduction can also be parallelized. But the overhead of the thread fork is high if N is small. Hence, we do not parallelize this part. However, this parallelization can adapt AT.

```
<1>  !$omp parallel
<2>  !$omp do PRIVATE(J_PTR,II,KK)
<3>        DO K=1,NUM_SMP
<4>          DO J_PTR=ISTART(K),IEND(K)
<5>            II = ICOL(J_PTR)
<6>            KK = YCOL(J_PTR)
<7>            YY(KK,K)=YY(KK,K)+VAL(J_PTR)*X(II)
<8>          END DO
<9>        ENDDO
<10> !$omp end do
<11> !$omp end parallel
<12>        DO K=1,NUM_SMP
<13>          DO I=1, N
<14>            Y(I) = Y(I) + YY(I,K)
<15>          ENDDO
<16>        ENDDO
```

**Fig. 1.** Outer loop parallelization with the COO-Column format.

## 3.2 Outer Loop Parallelization with COO-Row Format

If the sparse data is in the Row-wise COO format (COO-Row), parallel implementation with OpenMP is as shown in Fig. 2. The implementation of Fig. 2 is similar to the implementation of Fig. 1 except for YCOL and XCOL.

## 3.3 Outer Loop Parallelization with ELL-Row Format

If the sparse format is row-wise ELL (ELL-Row), we have two choices for parallelization. ELL has a bandwidth parameter, denoted NE. If NE is a large value, we can parallelize this loop. If NE is a small value, the parallelism is limited. If NE=2, the parallelism is only 2. In this situation, we can take the other parallelism for the N-loop, which is the number of rows in the sparse matrix. We do not decide the best method without considering the matrix specifications.

In this section, we show the parallelization for the N-loop, which is located in the inner loop with the ELL-Row format. Fig. 3 shows this implementation.

In Fig. 3, there is no reduction loop, which is an advantage of this format.

```
<1>!$omp parallel
<2>!$omp do PRIVATE(J_PTR,II,KK)
 <3>      DO K=1,NUM_SMP
 <4>          DO J_PTR=ISTART(K),IEND(K)
 <5>              II = ICOL(J_PTR)
 <6>              KK = XCOL(J_PTR)
 <7>              YY(II,K)=YY(II,K)+VAL(J_PTR)*X(KK)
 <8>          END DO
 <9>      ENDDO
<10>!$omp end do
<11>!$omp end parallel
<12>      DO K=1,NUM_SMP
<13>          DO I=1, N
<14>              Y(I) = Y(I) + YY(I,K)
<15>          ENDDO
<16>      ENDDO
```

**Fig. 2.** Outer loop parallelization with the COO-Row format.

```
 <1>      DO K=1,NE
 <2>          KK = N*(K-1)
 <3>!$omp parallel
 <4>!$omp do PRIVATE(J_PTR,II)
 <5>          DO I=1,N
 <6>              J_PTR = KK+I
 <7>              II = ICOL(J_PTR)
 <8>              Y(I)=Y(I)+VAL(J_PTR)*X(II)
 <9>          ENDDO
<10>!$omp end do
<11>!$omp end parallel
<12>      ENDDO
```

**Fig. 3.** Inner loop parallelization with the ELL-Row format.

## 3.4 Outer Loop Parallelization with the ELL-Row Format

Finally, we show the outer loop parallelization with the ELL-Row format in Fig. 4.

```
 <1>!$omp parallel
 <2>!$omp do PRIVATE(K,KK,I,J_PTR,II)
 <3>      DO J=1, NUM_SMP
 <4>          DO K=ISTART(J),IEND(J)
 <5>              KK = N*(K-1)
 <6>              DO I=1,N
 <7>                  J_PTR = KK+I
 <8>                  II = ICOL(J_PTR)
 <9>                  YY(I,J)=YY(I,J)+VAL(J_PTR)*X(II)
<10>              ENDDO
<11>          ENDDO
<12>      ENDDO
<13>!$omp end do
<14>!$omp end parallel
<15>      DO K=1,NUM_SMP
<16>          DO I=1,N
<17>              Y(I) = Y(I) + YY(I,K)
<18>          ENDDO
<19>      ENDDO
```

**Fig. 4.** Outer loop parallelization with the ELL-Row format.

## 4. Performance Evaluation

### 4.1 Machine Environment

We used the Plasma Simulator, which is the HITACHI SR16000/VL1 installed at the National Institute for Fusion Science. Each node contains 64 cores of the IBM POWER6 (5.0 GHz) microprocessor. We can execute 128 threads/node when we use simultaneous multithreading (SMT) execution.

The L1 cache is 64 KB/core, the L2 cache is 4 MB/core, and the L3 cache is 32 MB/2 cores. The memory on each node is 1024 GB. The OS is AIX 5L version 5.3. The theoretical peak is 1.280 TFLOPS/node. We used the HITACHI Fortran90 Compiler version V02-00-/B with the option "-opt=ss -omp."

We also used the Earth Simulator 2 (ES2) installed in the Japan Agency for Marine-Earth Science and Technology. Each node contains 8 cores of the NEC SX-9/E (3.2 GHz). The memory on the node is 128 GB. The theoretical peak is 819.2 GFLOPS/node. The inter-node connection is two levels of a fat tree, full-bisection connection. We used NEC SX Fortran90 Rev.404 2010/03/01 with the option "-Chopt."

We used OpenATLib version beta[7]. The sparse matrix format on OpenATLib is CRS, and its implementation of SpMV is also based on CRS. We used switch no. 11, which is the normal CRS implementation in OpenATI_DURMV provided by OpenATLib.

### 4.2 Test Matrices

For sample matrices, we took matrices from The University of Florida Sparse Matrix Collection[8]. The matrix information and the mean, derivation, and $D_{mat}$ defined in (4) are shown in Table 1. In matrix no. 3, which is torso1, the overflow memory space is in the ELL format in our experiment. Hence, we removed the data of the ELL format.

**Table 1.** The Test Matrices: All matrices are unsymmetric.

(a) Test Matrices: Set I

| No. | Name | N | NNZ | Mean $\mu$ | Derivation $\sigma$ | $D_{mat}$ | Field |
|---|---|---|---|---|---|---|---|
| 1 | chipcool0 | 20082 | 281150 | 14.00 | 2.69 | 0.19 | 2D/3D |
| 2 | chem_master1 | 40401 | 201201 | 4.98 | 0.14 | 0.02 | |
| 3 | torso1 | 116158 | 8516500 | 73.31 | 419.58 | 5.72 | |
| 4 | torso2 | 115067 | 1033473 | 8.91 | 0.58 | 0.06 | |
| 5 | torso3 | 259156 | 4429042 | 17.09 | 4.39 | 0.25 | |
| 6 | memplus | 17758 | 126150 | 7.10 | 22.03 | 3.10 | Electric circuit |
| 7 | ex19 | 12005 | 259879 | 21.64 | 12.28 | 0.56 | Fluid dynamics |
| 8 | poisson3Da | 13514 | 352762 | 26.10 | 13.76 | 0.52 | |
| 9 | poisson3Db | 85623 | 2374949 | 27.73 | 14.71 | 0.53 | |
| 10 | airfoil_2d | 14214 | 259688 | 18.26 | 3.94 | 0.21 | |
| 11 | viscoplastic2 | 32769 | 381326 | 11.63 | 13.95 | 1.19 | Materials |

(b) Test Matrices: Set II

| No. | Name | N | NNZ | Mean $\mu$ | Derivation $\sigma$ | $D_{mat}$ | Field |
|---|---|---|---|---|---|---|---|
| 12 | xenon1 | 48600 | 1181120 | 24.30 | 4.25 | 0.17 | Materials |
| 13 | xenon2 | 157464 | 3866688 | 24.55 | 4.06 | 0.16 | |
| 14 | wang3 | 26064 | 177168 | 6.79 | 0.43 | 0.06 | Semiconductor device |
| 15 | wang4 | 26068 | 177016 | 6.79 | 0.43 | 0.06 | |
| 16 | ecl32 | 51993 | 380415 | 7.31 | 3.35 | 0.45 | |
| 17 | sme3Da | 12504 | 874887 | 69.96 | 34.92 | 0.49 | Structural |
| 18 | sme3Db | 29067 | 2081063 | 71.59 | 37.06 | 0.51 | |
| 19 | sme3Dc | 42930 | 3148656 | 73.34 | 36.98 | 0.50 | |
| 20 | epb1 | 14734 | 95053 | 6.45 | 0.57 | 0.08 | Thermal |
| 21 | epb2 | 25228 | 175027 | 6.93 | 6.38 | 0.92 | |
| 22 | epb3 | 84617 | 463625 | 5.47 | 0.54 | 0.10 | |

### 4.3 Speed up by using COO and ELL formats

Fig. 5 shows $SP_{crs/ell}$, which is the speedup ratio for SpMV in CRS in the HITACHI SR16000/VL1 from 1 thread to 128 threads. Fig. 5 indicates the following:

1. Speedup is mainly observed in 1 thread.

2.  ELL is more efficient than COO when the number of threads is low, such as 1 or 4. Matrix no. 6, which is memplus, is the exception of the ELL format.

3.  According to the increase of the number of threads, ELL-Row is the best. However, there is no advantage of ELL for 64 and 128 threads.

As a result, the maximum 2.45x speedup in 1 thread in matrix no. 2, chem_master1, with ELL-Row Inner-Parallelized, is obtained.

2.  According to the increase of the number of threads, ELL-Row outer-parallelized is the best.

A 151x speedup with ELL-Row inner-parallelized is obtained in no. 2, which is chem_master1.

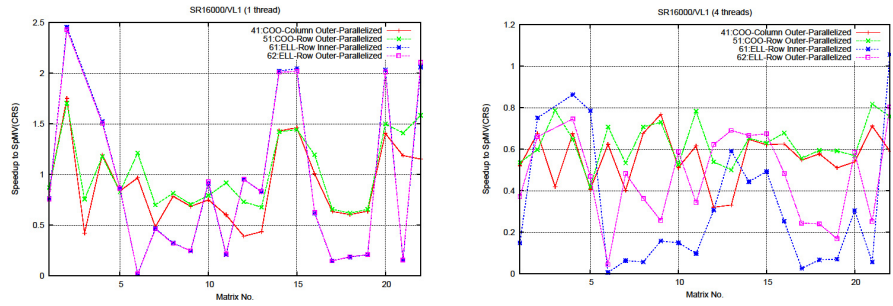

(a) 1 Thread                    (b) 4 Threads

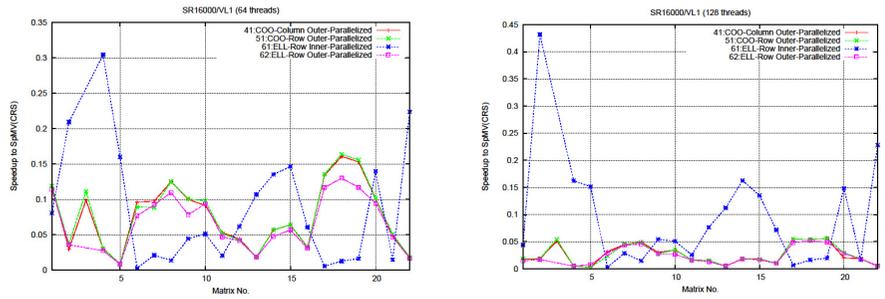

(c) 64 Threads                    (d) 128 Threads

**Fig. 5.** $SP_{crs/ell}$ in the HITACHI SR16000/VL1.

Fig. 6 shows $SP_{crs/ell}$ in the ES2. The conclusions of Fig. 6 are as follows:

1.  More than 100x speedups are established by using ELL. The exception is no. 6, memplus. In this case, the COO-Row format is the best. The speedup is 2.75 x.

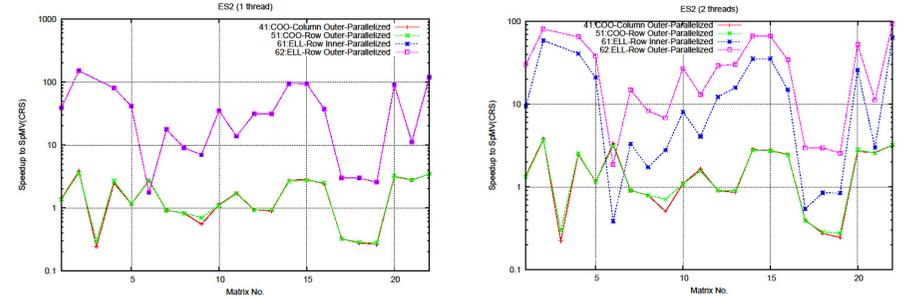

(a) 1 Thread                    (b) 2 Threads

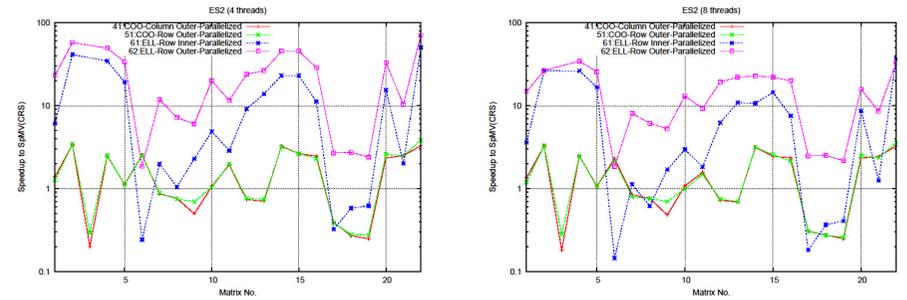

(c) 4 Threads                    (d) 8 Threads

**Fig. 6.** $SP_{crs/ell}$ in the ES2.

### 4.4  Transformation Overheads and The $D_{mat}{}^i$ - $R_{ell}{}^i$ Graph

Fig. 7 shows $TT_{ell}{}^i$ in one thread in the HITACHI SR16000/VL1 and the ES2. $TT_{ell}{}^i$

indicates the data transformation overheads based on one time of SpMV with CRS.

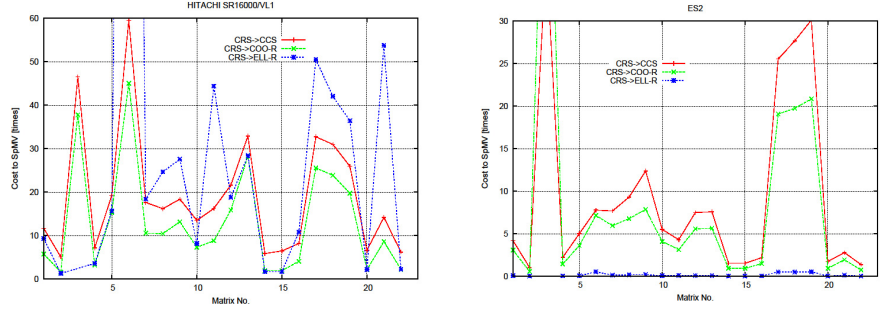

(a) HITACHI SR16000/VL1

(b) ES2

**Fig. 7.** $TT_{ell}$ in one thread of HITACHI SR16000/VL1 and ES2.

In Fig. 7, some matrices require high overhead transformation that is more than 20x–50x. Especially, the overhead is high for matrices no. 6 and nos. 17–19. In contrast, in the ES2, the ELL overheads are very low, such as 0.01x–0.51x.

Fig. 8 shows the $D_{mat}{}^i$ - $R_{ell}{}^i$ graph. As explained in Section 2.2, a value higher than 1.0 in the y-axis is the threshold for efficiency in the run-time data transformation.

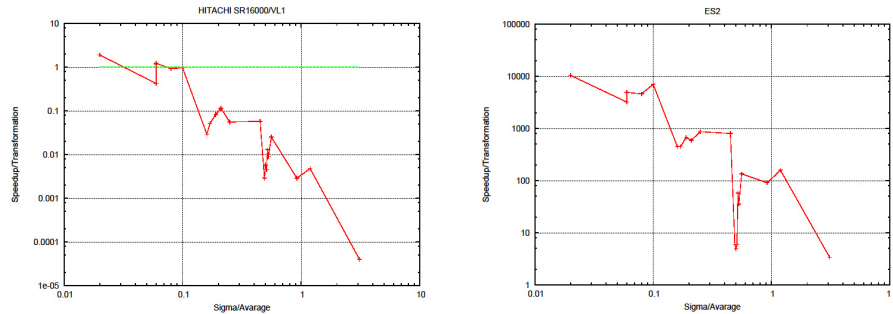

(a) HITACH SR16000/VL1

(b) ES2

**Fig. 8.** The $D_{mat}{}^i$ - $R_{ell}{}^i$ graph outer-parallelized with ELL-Row in one thread.

Fig. 8 includes the following important conclusions:
- All matrices with $D_{mat}$ from 0.02 to 3.10 are an advantage in the ES2.
- Matrices with $D_{mat}$ less than 0.1 are an advantage in the SR16000/VL1.

Please note that computing $D_{mat}$ requires a very low cost. This is a good example to show the effectiveness of the $D_{mat}{}^i$ - $R_{ell}{}^i$ graph to use run-time AT for sparse data transformation.

### 4.5 Discussion

Fig. 8 shows that matrices with large $D_{mat}$ values cause high transformation overheads. This occurs because, to change to the ELL format from the CRS format, the costs of finding the nonzero positions and the inclusion of zero values are increased if the matrix has non-constant bandwidth. ELL is compact if the matrix forms a perfect band. In this situation, the $D_{mat}$ value is very small because the derivation goes close to zero. Moreover, if the matrix has a perfect band, no zero element is filled and, hence, there is no memory access. This causes very efficient computation for the ELL format with non-zero elements.

The results show that the $D_{mat}{}^i$ - $R_{ell}{}^i$ graph can be used for the decision of run-time AT for sparse matrix data transformation.

## 5. Conclusion

In this paper, we evaluated the possibility of run-time sparse matrix data transformation from the CRS format to the COO and the ELL formats. We proposed an AT method using the $D_{mat}{}^i$ - $R_{ell}{}^i$ graph, which plots the derivation/average for the number of non-zero elements per row, which is described as $D_{mat}{}^i$ on the X-axis, and the ratio of (SpMV speedups)/(transformation time to ELL) to the CRS format, which is described as $R_{ell}{}^i$ on the Y-axis.

The experimental results show that the ELL format is very effective in the Earth Simulator 2; for example, a 151x speedup is established with the inner-parallelized ELL-Row. In addition, the transformation overhead is also very low, such as 0.01 to 1.0 SpMV with CRS. The $D_{mat}{}^i$ - $R_{ell}{}^i$ graph can be well modeled for the effectiveness of transformation based on the derivation/average value. Hence, we show the possibility of using the graph to run-time AT.

In this study, we do not show the parallel implementations of the data transformation processes from the CRS format to the COO and the ELL formats. Evaluation with parallelized transformations and the evaluation of $D_{mat}{}^i$ - $R_{ell}{}^i$ graphs are future work. In addition, evaluating the transformation to other formats, such as BCSR, which enables cache blocking, is important future work.

**Acknowledgments** This work is partially supported by a Grant-in-Aid for Scientific Research (B) "Development of the Framework to Support Large-scale Numerical Simulation on Multi-platform," No. 21300017, and a Grant-in-Aid for Scientific Research (B) "Development of Auto-tuning Specification Language Towards Manycore and Massively Parallel Processing Era," No. 21300007. For the use of Earth Simulator 2, we sincerely thank those associated with the Earth Simulator Collaboration Projects, FY2010, "Development of Adaptive High Accuracy Libraries" at the Japan Agency for Marine-Earth Science and Technology.